\documentclass[aps,twocolumn,showpacs,preprintnumbers]{revtex4}
\usepackage{graphicx}
\usepackage{psfig}
\usepackage{dcolumn}
\usepackage{bm}

\begin{document}

\preprint{submitted to PRL}

\title{Elastic Energy, Fluctuations and Temperature for Granular Materials}

\author{Lou Kondic$^{(1)}$ and R. P. Behringer$^{(2)}$}
\affiliation{$^{(1)}$Department of Mathematical Sciences 
\& Center for Applied Mathematics \& Statistics\\ 
New Jersey Institute of Technology, Newark, NJ 07102\\ } 
\affiliation{$^{(2)}$Department of Physics \& Center for Nonlinear and
Complex Systems\\ Duke University, Durham NC, 27708-0305 }

\date{July 1, 2003}
\begin{abstract}
We probe, using a model system, elastic and kinetic energies for
sheared granular materials.  For large enough $P/E_y$
(pressure/Young's modulus) and $P/\rho v^2$ ($P/$kinetic energy
density) elastic dominates kinetic energy, and energy fluctuations
become primarily elastic in nature.  This regime has likely been
reached in recent experiments.  We consider a generalization of the
granular temperature, $T_g$, with both kinetic and elastic terms and
that changes smoothly from one regime to the other.  This $T_g$ is
roughly consistent with a temperature adapted from equilibrium
statistical mechanics.

\end{abstract}

\maketitle We explore the role of elasticity in the energy and energy
fluctuations of sheared dense granular systems.  For dilated gas-like
granular states, energy fluctuations are frequently described in terms
of a temperature, defined as the fluctuating part of the kinetic
energy, $T_k \equiv m< v^2 >/2$.  Here, $v$ is the local random
component of the velocity.  This definition is predicated on
assumptions such as molecular chaos, absence of correlations, and
short-lived collisions, that do not always apply.  For dense systems,
a very different concept, Edwards entropy, has been
proposed~\cite{edwards}.  This quantity is the logarithm of the number
of jammed configurations consistent with all constraints on the
system, and the Edwards temperature is $T_E^{-1} = \partial
S_E/\partial V$, where $V$ is the system volume.

Both of these pictures assume that minimal energy is stored in
compressional modes of the particles.  This assumption is valid when
the pressure is small compared to the Young's modulus, $E_y$.
However, there are situations when this need not be the case.  The
main goal of this letter is to analyze via discrete element
simulations (DES) the storage of energy and energy fluctuations for
dense granular material subject to shearing. 

In order to better establish a context, we estimate the relative
importance of kinetic and elastic energy for a simplified system.  We
imagine that spherical particles with a typical velocity, $v$, are
subject to an applied force, $L$, at each contact.  For simplicity,
assume at first just a pair of opposing contacts, so that there is an
effective pressure, $P= L/A$, where $A = \pi R^2$, and $R$ is the
radius of a sphere.  The elastic energy per contact (assuming a
Hertz-Mindlin contact law) is $\epsilon \propto L^{5/3}.$ The ratio of
elastic to kinetic energy for a particle is $R_E = \epsilon/K = C
(P/E_y)^{2/3} (P/\rho v^2)$, where $C$ is an $O(1)$ constant that
depends on the Poisson ratio, $E_y$ is the Young's modulus, and $\rho$
is the bulk density. By multiplying by the number of contacts per
particle, it is possible to generalize this to a more realistic
situation.  When $R_E$ is small, elastic energy is irrelevant, and
{\em vice versa} when it is large, it should be included in a
description of the system.  

We now consider $R_E$ for two representative cases.  For the
experiments of~\cite{howell_99}, $E_y = 5 MPa$, typical $P\sim 160
Pa$, and $\rho=1.2$ g/cm$^3$.  Typical speeds ranged over $6 \times
10^{-4} m/s \leq v \leq 2\times 10^{-2} m/s$, so that $0.3 \leq R_E
\leq 4 \times 10^2$.  For glass spheres~\cite{losert_00}, $E_y$ is
larger by roughly a factor of $5000$.  Assuming densities $\rho \sim
2$ g/cm$^3$, and pressures corresponding to the base of a column $10$
cm high in a gravitational field, $R_E \simeq 1$ for velocities of a
few mm/s.

Hence for slowly sheared dense granular systems, there exist velocity
regimes for which elastic energy is the dominant mode of energy
storage.  In such a setting, neither $T_k$ nor $T_E$ is likely to
provide a good measure of the random nature of the system.

In this context, we propose an extension of granular ``temperature''
that contains information on fluctuations of the elastic energy, and
then compare this extension to a relation for temperature drawn from
statistical mechanics.  (A somewhat similar approach of applying
equilibrium statistical theory has been taken in recent works on
foams~\cite{ono_02} and granular systems~\cite{makse_02}.)  To carry
out this exploration, we use DES of 2D particles that are subject to
plane shear and (possibly) compression.

The generalization of ``temperature'' that we consider is based on the
classical idea that for a lattice of elastic particles, the average
fluctuating energy/particle is $3k_B T$.  Using this as a heuristic
guide, we define a generalized temperature that is roughly $T_g = m
<v^2>/2 + k <x^2>/2$, where $v$ corresponds to the fluctuating part of
the velocity, and $x$ to the fluctuating part of the compression of a
particle.  Note that this definition provides a simple bridge between
the extremes of a gas-like state and a highly compressed slowly
evolving state.

\begin{figure}[t]
\vfill\centerline{\fbox{\psfig{figure=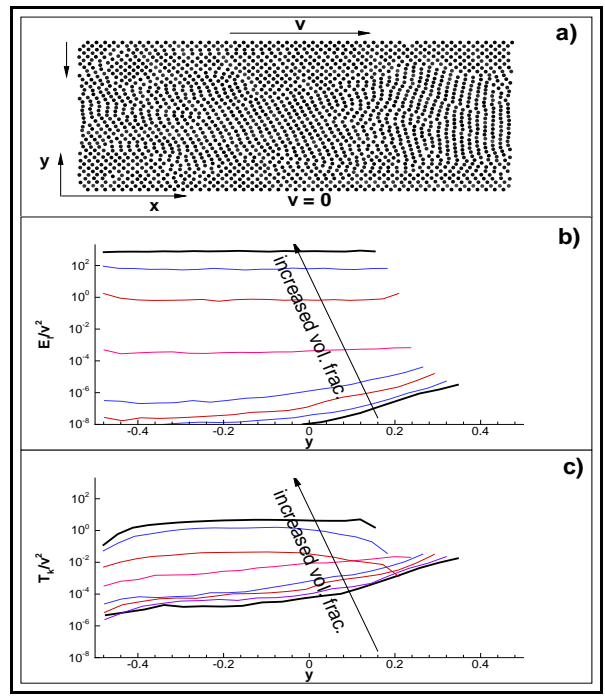,height=3.5in,width=3in}}}
\caption{(a) The geometry of the simulations
(only centers of the particles are shown). (b,c,)
Elastic energy and kinetic temperature, scaled by $v^2$, as functions of 
the distance from the shearing wall, $y$.  
The shearing wall is to the right, and the shearing is in 
the out-of-plane direction ($v=0.1$). 
} 
\label{fig:ener_temp_var}
\end{figure}

In the simulation, particles are confined between two inpenetrable
straight parallel boundaries, as sketched in
Fig.~\ref{fig:ener_temp_var}a.  The top boundary, which is 50 mean
particle diameters ($d_m$) long, moves at a steady speed, and induces 
shearing in the system.  The boundary conditions in the shearing direction are
periodic.  This system avoids the nonuniformity that characterizes
physical experiment, which typically exhibit shear bands~\cite{howell_99}.  
Although we could have chosen to use the even simpler Lees-Edwards
conditions, we have concentrated on the present model
so as to explore the influence of boundaries, typically present in physical
experiments.

These simulations closely follow the soft-disk/sphere model to
describe various granular systems (see \cite{kondic_99} and references
therein).  Here, we concentrate on two-dimensional polydisperse disks
in a zero-$g$ environment.  The walls are made of identical particles
that are rigidly attached.  Forces between the particles have a normal
component given by $ {\bf F}_N = \left[ k_f x - \gamma_N \bar m ({\bf
v}_{i,j} \cdot {\bf\hat n})\right]$ where $k_f$ is a force constant,
$r_{i,j} = |{\bf r}_{i,j}|$, ${\bf r}_{i,j} = {\bf r}_i - {\bf r}_j$,
${\bf \hat n} = {{\bf r}_{i,j}/ r_{i,j}}$, $d = {(d_i + d_j)/2}$,
$d_{i,j}$ are the diameters of the particles $i$ and $j$, $x =
d-r_{i,j}$ is the compression, ${\bf v}_{i,j} = {\bf v}_i - {\bf
v}_j$, $\bar m$ is the reduced mass, and $\gamma_n$ is the damping
constant related to the coefficient of restitution, $e_n$.  The
parameters represent photoelastic disks~\cite{howell_99}; in
particular, $\gamma_n$ corresponds to $e_n = 0.5$.  The tangential
force is given by ${\bf F}_S = sign(-v_{rel}^t) min \left ( \gamma_s
\bar m |v_{rel}^t|, \nu_k |{\bf F}_N^c |\right ) {\bf\hat s}$ where
$v_{rel}^{t}$ is the relative velocity in the tangential direction
${\bf\hat s}$, $\gamma_s = {\gamma_n/2}$ and $\mu_k$ is the
coefficient of friction between the particles.  The equations of
motion are then integrated using a $4$th order predictor-corrector
method.  Additional simulations (to be presented
elsewhere) show that variation of the parameters or the force model
modify only details of the results.

The simulations are performed with approximately $2000$ polydisperse
particles, with a radius variability of $10$\%.  While polydispersity
is important to avoid crystallization, the details of the
size distribution are not: the results are very similar for different
ranges of particle sizes, or a bidisperse distribution.  Particles are
initially placed on a lattice, given random velocities, and the system
is then very slowly compressed to a desired volume fraction, $\nu$.
The results that follow use $t$, the time it takes the shearing wall
to travel once across the domain, as a time scale, and $l/t$ ($l=50
d_m$) as a velocity scale.

The quantities below are calculated using space-time averaging.  
Thus, the system is divided into cells, and averaged quantities are calculated
for each cell.  In particular, the kinetic temperature is defined by 
\[
T_k= {1\over 2} \left[ \langle m(u')^2 \rangle + \langle m(v')^2
\rangle + {\beta\over 4} \langle m (d_p\omega')^2 \rangle \right]\, ,
\]
where $u,v$ are the components of particle velocity, $d_p$ 
is the diameter of a particle, $m$ is its mass ($m\sim d_p^2$),
$\beta = 1/4$ for disks, and $\omega$ is the angular velocity.  The primed
averages are defined with zero mean, e.g, $\langle u'v'\rangle =
\langle uv\rangle - \langle u\rangle \langle v\rangle$.  

The elastic energy is obtained by averaging {\it per collision}, not
per particle.  The difference between the two is significant for dense
granular systems considered here, since particles typically experience
multiple collisions.  If $x_{j,c}$ is the compression of particle $j$
due to the collision $c$, then our definition of the elastic energy in
cell $l$ is
\begin{equation} 
\label{eq:elasw}
E_{e,l} = {1\over N_t n_l}{k_f\over 2} \sum_{k=1}^{N_t}
\sum_{j=1}^{n_{l}}\sum_{c=1}^{n_{c,j}}  \left[ x_{j,c}\right]^2,
\end{equation} 
where $n_l$ is the number of particles in cell $l$ at a given time,
and $\bar n_l$ is the average number of particles during the period of
$N_t\gg 1$ time steps (in practice, the averaging time scale is 
sufficiently short so that to a high degree of accuracy $n_l = \bar n_l$).  
Definition~(\ref{eq:elasw}) ignores the energy used to overcome friction, and
its form clearly depends on the form of the normal force, e.g., 
for a 3D systems, for which a Hertzian interaction law 
($F_N\sim x^{3/2}$) is more appropriate, $E_{e,l}$ is
of different form as well.

\begin{figure}
\vfill\centerline{\fbox{\psfig{figure=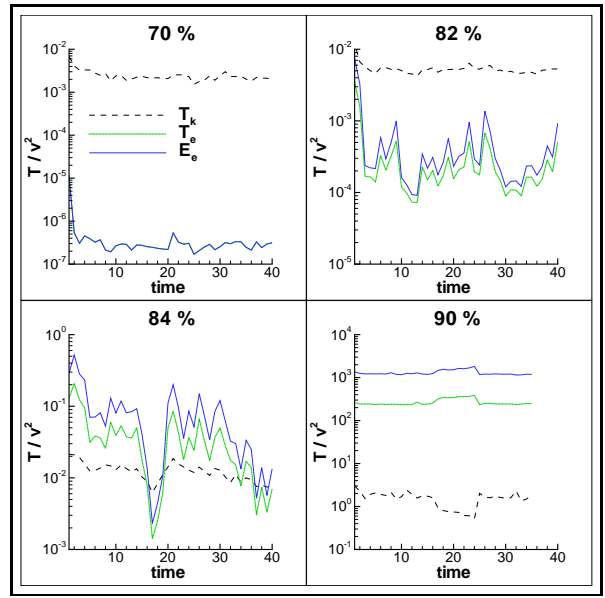,height=3in,width=3in}}}
\caption{Elastic energy, kinetic and elastic temperatures
for various $\nu$'s, scaled by $v^2$.  
}
\label{fig:ener_temp_time}
\end{figure}

Figure~\ref{fig:ener_temp_var} shows the elastic energy and kinetic
temperature (scaled by the square of the shearing velocity $v$, and by
the average mass of a particle) vs. distance, $y$, from the shearing
wall.  For these simulations, $\nu$ is continuously increased by
(slow) compression.  $\nu$ increases from $65$\% (bottom) to $90$\%
(top) (note that in 2D, random close packing and cubic close packing
correspond to about $85$\% and ${2\pi/\sqrt{3}}\approx 90$\%,
respectively).  Clearly, as $\nu$ is increased, there is a transition
region (about $\nu_c=80$\%) where the energy stored in the internal
degrees of freedom (elastic energy) becomes more relevant than the
kinetic energy (see also Fig.~\ref{fig:ener_temp_time}).  
The $y$-dependence of the results is rather weak and
becomes even weaker for higher $\nu$'s.  Hence, hereafter, we ignore
the $y$-dependence and use system averages of locally computed
quantities.  

As $\nu$ increases, the energy is mainly elastic, and $T_k$ loses its
relevance.  In order to have a quantity that might play the same role
as $T_k$ in a dense granular system we propose a generalized granular
'temperature' by
\begin{equation}
T_g = T_k + T_e,
\end{equation}
as a sum of $T_k$, and the `elastic' part $T_e$.  $T_e$ is defined as
the mean fluctuation of {\it elastic} energy, in a manner similar
to $T_k$, which is the mean fluctuation of {\it kinetic} energy.  This
definition follows the classical statistical mechanics result where
the mean fluctuations in the combined elastic and kinetic energy of an
oscillator are proportional to the temperature.  However, unlike this
classical case, there is no reason to expect equipartition between
elastic and kinetic modes.  Rather, the ratio $T_e/(T_e + T_k)$ varies
from 0 in the dilute limit, to 1 in the dense limit.

The definition of $T_e$ requires some care due to multiple collisions.
One simple and natural definition that is consistent with the classical
statistical definition of temperature for an oscillator is as follows.
We first define the average elastic energy per particle in cell $l$ as
\begin{equation}\label{eq:ave_ener}
\langle E_{e,l}\rangle = {k_f\over 2}n_c \langle x_{l}\rangle^2 =
{k_f\over 2}n_c 
\left[ {1\over N_t \bar n_l n_c}\sum_{k=1}^{N_t}
\sum_{j=1}^{n_{l}}\sum_{c=1}^{n_{c,j}}  x_{j,c}\right]^2\, ,
\end{equation}
where $\langle x_l\rangle$ is the average compression per collision, and $n_c$ is
the average number of collisions per particle.  Then, 
\begin{equation} \label{eq:tempdef}
T_{e,l} = {k_f\over2}n_c \langle\delta x^2\rangle = 
{k_f\over2}n_c \langle (x_{j,c} - \langle x_l\rangle)^2\rangle =
E_{e,l} - \langle E_{e,l}\rangle
\, ,
\end{equation}
where the last equality easily follows using~(\ref{eq:elasw}) and
(\ref{eq:ave_ener}).

Figure~\ref{fig:ener_temp_time} shows $T_k$, $T_e$, and $\langle E_{e}\rangle$
vs. time, for four different $\nu$'s.  Unlike the results of
Fig.~\ref{fig:ener_temp_var}, these results are obtained after
shearing for long times at fixed $\nu$'s.  For higher $\nu$'s, clearly
$T_e \gg T_k$.  Interestingly, there are rather large fluctuations of
the results with $\nu$ just above $\nu_c\approx 80$\%, corresponding
to the regime where elastic energy becomes predominant.  Also,
experiments have indicated a phase transition for comparable densities
which may be related~\cite{howell_99} (see also~\cite{campbell_02}).  
Generally, one might expect
both glassy and/or jamming phenomena to dominate this regime, a point
that we will explore elsewhere.

We further interpret $T_g$ by comparing it to an alternative
definition from statistical mechanics.  We consider, among various
possibilities, the standard relation~\cite{mcquarrie}
\begin{equation} {d U\over d T } = {\delta U^2 \over T^2}\, , 
\label{eq:model}
\end{equation}
where $U$ is the total energy in the (usually conservative) system,
$T$ is the temperature, and $\delta U^2 = \langle U^2 \rangle -
\langle U \rangle^2$.  We use $U = E_{kin} + \langle E_{e}\rangle$,
and ask whether it makes sense to define $T = T_g = T_k + T_e$.  To
check this idea, we now think of~(\ref{eq:model}) as a defining
equation for $T_m= {\sqrt{\delta U^2}/({dU/dT_g})}$ ( here $T_m$ stands for 
`model temperature').  The agreement
between $T_m$ and $T_g$ will provide some indication of the utility of
the definition for $T_g$.  Note that we should not expect perfect or
possibly even any agreement, since we are considering a strongly
dissipative system far from equilibrium.  We emphasize
that there are no fitting parameters and that the results that follow
are obtained directly from simulations.

\begin{figure}[htb]
\vfill\centerline{\fbox{\psfig{figure=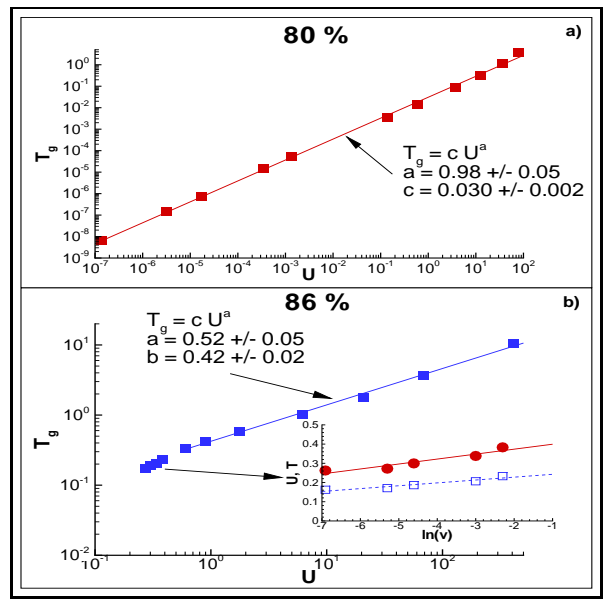,height=3in,width=3in}}}
\caption{Generalized temperature versus average energy for two different $\nu$'s
(increased $U$ corresponds to increased shearing velocity $v$).
The lines are the least square fits to the data.  In the inset
of b), $U$ (solid line, filled circles) and $T_g$ (broken line, squares) are 
plotted versus $ln(v)$ for slow shearing. 
}
\label{fig:tempfit}
\end{figure}
\clearpage
\begin{figure}[thb]
\vfill\centerline{\fbox{\psfig{figure=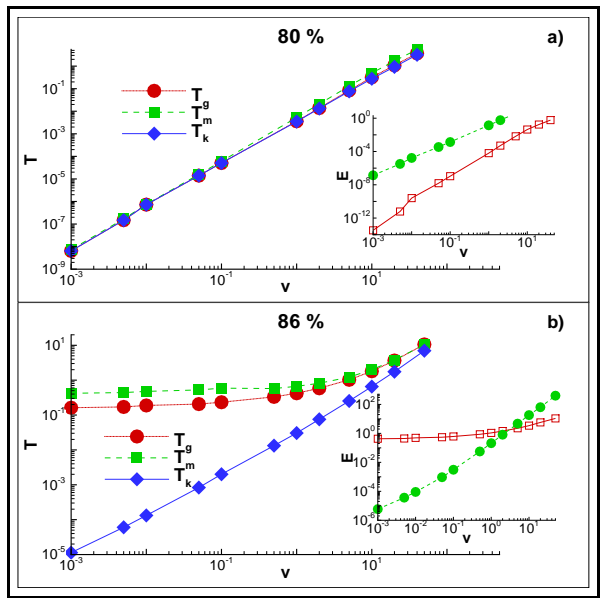,height=3in,width=3in}}}
\caption{Generalized, kinetic, and `model' temperature (see the text and Eq.~(\ref{eq:model}))
for two volume fractions.
Error bars (resulting from statistical uncertainty of the results, as well as from the 
uncertainty introduced by the least square fits from Fig.~\ref{fig:tempfit})
are approximately represented by the size of the symbols.  The insets show kinetic
energy (filled circles) and average elastic energy (squares).
}
\label{fig:model}
\end{figure}

Figure~\ref{fig:tempfit} shows typical data used to determine $c_v =
{dU/dT_g}$ needed for $T_m$.  The range of $U$'s shown in
Fig.~\ref{fig:tempfit} corresponds to $ 0.0001 < v < 40$.  For
$\nu=80$\%, $U \sim T_g$ over almost 10 decades.  For $\nu = 86$\%, $U
\sim T_g^a$, where $a \simeq 2.0$.  However, in this case, there is a
deviation from the power law fit for slow shearing, and we are limited
to a smaller range of $U$'s, since most of the energy is stored in
the system as elastic energy.  In order to obtain better data for
$U(T_g)$ for slow shearing at higher densities, we determine
$U$ and $T_g$ as functions of $v$, see Fig.~\ref{fig:tempfit}b.  We
then find to a good approximation that both $U$ and $T_g$, are
proportional to $ln(v)$.  This logarithmic dependence for slow
shearing is, to the best of our knowledge, the first computational
confirmation of recent experimental results~\cite{hartley_03}, and will
be presented in more detail elsewhere. For our purposes here, it is
sufficient to extract $U(T_g)$.  

The dependence of $U(T_g)$ is interesting.  We note that $U \sim T_g$
for $\nu = 80$\% is similar to a recent simulation of sheared
foams~\cite{ono_02}.  However, the result $U\sim T_g^2$ for $\nu =
86$\% and for not too slow shearing 
($U> 1$, see Fig.~\ref{fig:tempfit}b) is striking and deserves some comment.
Although the increased role of elastic energy is important here, this
does not appear to be the whole story, as seen in the context of
Fig.~\ref{fig:model}.  Here, we show that even for this high $\nu$,
for fast enough shearing the kinetic energy is still dominant.  Our
preliminary interpretation is that increased volume fraction
contributes significantly to decreased mobility (jamming) of the
granular system, therefore reducing the increase of $T_g$ with $U$.

Figure~\ref{fig:model} contains a summary of the various types of
temperatures considered in this study for two $\nu$'s.  As already
noted, for $\nu=80$\%, most of the energy is still kinetic, while for
$\nu=86$\%, elastic energy is essential.  For $\nu=80$\%, $T_k$ is
dominant, and also it satisfies the model~(\ref{eq:model}) since $T_k
\approx T_m$.  However, for $\nu =
86$\% and for slow shearing, $T_k$ is smaller by $4$ orders of
magnitude than $T_g$ and $T_m$.  Thus, $T_k$ cannot be used to even
approximately describe a dense slowly sheared granular system.  This
difference decreases for higher shearing rates, but only at very high
shearing is there good agreement.

The agreement between $T_g$ and $T_m$ is not perfect, although prefect
agreement is not to be expected. These studies demonstrate the clear
need to incorporate elastic energy and elastic fluctuations, and that
$T_g$ has utility as a generalized granular temperature. Nevertheless,
there remain many open questions regarding the extent to which the
various temperatures serve similar functions to their molecular
counterpart.  We will present detailed results addressing this type of
question elsewhere.  Here we note that at least qualitatively one can
show that these temperatures can be used in the context of thermal
conduction, i.e.~the there is a flow of heat from hot to cold regions.
We also note that the distributions of both kinetic and elastic energy
are strongly non-Gaussian for a significant range of the parameters
analyzed here.

There are other possible tests of the proposed generalization, as
recently discussed on the context of sheared foams~\cite{ono_02}.  The
ultimate test will be to perform physical experiments where
the validity of the proposed concepts can be verified.  
In the theoretical direction, it will be of interest to relate
the generalized granular temperature proposed here to the one
resulting from Edwards-entropy based approach.  

We acknowledge support by NASA NAG3-2367 and NAG3-2372.  RPB acknowledges
support by NSF grants DMR-0137119, DMS-0204677 and DMS-0244492.

\end{document}